\documentclass[aps,pre,amsmath,amssymb,amsfonts,lengthcheck,twocolumn,superscriptaddress]{revtex4-2}
\usepackage{graphicx}
\usepackage{subfigure}
\usepackage{amsthm}
\usepackage{verbatim}
\usepackage{dcolumn}
\usepackage{bm}
\usepackage{epsf}
\usepackage{color}
\usepackage[colorlinks=true,citecolor=blue,linkcolor=blue,urlcolor=blue]{hyperref}%
\usepackage{xcolor}
\usepackage{dsfont}
\usepackage{tikz}
\usepackage{todonotes}
\usepackage{multirow}
\usepackage{xfrac}
\usepackage{makecell}
\usepackage[T1]{fontenc}

\newcommand{\pd}{\partial}

\newcommand{\bla}{bla\\bla\\bla\\bla\\bla}

\newcommand{\mrm}[1]{\mathrm{#1}}

\DeclareMathAlphabet\mathbfcal{OMS}{cmsy}{b}{n}

\makeatletter
\newcommand{\currentfontsize}{\f@size pt}
\makeatother

\makeatletter
\newcommand\footnoteref[1]{\protected@xdef\@thefnmark{\ref{#1}}\@footnotemark}
\makeatother

\begin{document}

\title{Endoreversible Stirling cycles: plasma engines at maximal power}

\author{Gregory Behrendt}
\email{gregorb2@umbc.edu}
\affiliation{Department of Physics, University of Maryland, Baltimore County, Baltimore, MD 21250, USA}
\author{Sebastian Deffner}
\affiliation{Department of Physics, University of Maryland, Baltimore County, Baltimore, MD 21250, USA}
\affiliation{Quantum Science Institute, University of Maryland, Baltimore County, MD 21250, USA}
\affiliation{National Quantum Laboratory, College Park, MD 20740, USA}



\begin{abstract}
Endoreversible engine cycles are a cornerstone of finite-time thermodynamics. We show that endoreversible Stirling engines operating with a one-component plasma as working medium run at maximal power output with the Curzon-Ahlborn efficiency. As a main result, we elucidate that this is actually a consequence of the fact that the caloric equation of state depends only linearly on temperature and only additively on volume. In particular, neither the exact form of the mechanical equation of state, nor the full fundamental relation are required. Thus, our findings immediately generalize to a larger class of working plasmas, far beyond simple ideal gases. In addition, we show that for plasmas described by the photonic equation of state the efficiency is significantly lower. This is in stark contrast to endoreversible Otto cycles, for which photonic engines have an efficiency larger than the Curzon-Ahlborn efficiency.
\end{abstract}

\maketitle


\section{Introduction}

Among all the various states of matter, plasmas have a unique place. In particular, quantum plasmas exist only in the most extreme environments, such as the interior of stars, in the early universe, or more mundanely in highly intense laser fields \cite{Conde2020}. In its simplest form, a plasma is a superheated gas, in which electrons have been stripped from atoms, creating a mixture of positively charged ions and negatively charged electrons \cite{Conde2020}. It provides the circumstances for nuclear fusion, which is a nuclear reaction where two light atomic nuclei combine to form a heavier one, releasing a tremendous amount of energy \cite{morse2018nuclear}. The potential technological applications of nuclear fusion are tremendous, which is why  its realization is considered one of the ``14 Grand Challenges for Engineering in the 21st Century'' \cite{fusion}. 

From a thermodynamic point of view, also nuclear reactions in plasmas can be understood as heat engine cycles \cite{Segantin2020FED,Terahara2023FED}. In fact, the electrostatic interactions within a plasma behave very akin to the usual mechanical pressure in typical gases \cite{Avinash2010PP}. Interestingly, the somewhat natural engine cycle for plasma engines is the Stirling cycle \cite{Avinash2014PINSA}. The question then arises what the efficiency of such an engines is at maximal power output.

In the present work, we answer this question within the framework of endoreversible thermodynamics \cite{Hoffmann1997}. In endoreversible thermodynamics one assumes that all processes are slow enough that the system \emph{locally equilibrates}, yet the processes are too fast for the system to reach a state of equilibrium with the environment. More specifically, imagine an engine, whose working medium is at equilibrium at temperature $T$. However, $T$ is not equal to the temperature of the heat bath, $T_\mrm{bath}$, and thus there is a temperature gradient at the boundaries of the engine. Now further imagine that the engine undergoes a slow, cyclic state transformation, where slow means that the working medium remains \emph{locally} in equilibrium at all times. Then, from the point of view of the environment the device undergoes an irreversible cycle. Such state transformations are called \emph{endoreversible} \cite{Hoffmann1997}, which means that locally the transformation is reversible, but globally irreversible. 

In a seminal work, Curzon and Ahlborn showed \cite{Curzon1975AJP} that the efficiency of a Carnot engine undergoing an endoreversible cycle at maximal power is given by,
\begin{equation}
\label{eq:CA}
\eta_\mrm{CA}=1-\sqrt{\frac{T_c}{T_h}}\,,
\end{equation}
where $T_c$ and $T_h$ are the temperatures of the cold and hot reservoirs, respectively. Since its discovery the Curzon-Ahlborn efficiency \eqref{eq:CA} has received a great deal of attention.  For instance, it has been found that also endorversible Otto \cite{Deffner2018Entropy} and Brayton \cite{Ferketic2023EPL} engines operating with ideal gases have the same efficiency. However, it has also been shown that whether or not a finite time Carnot cycle really assumes $\eta_\mrm{CA}$ is determined by the ``symmetry'' of the dissipation \cite{Esposito2010} and on the specific form of the fundamental relation \cite{Leff1987,Rezek2006,Abah2012,Deffner2019book,Kloc2019,Myers2020PRE,Smith2020JNET,Myers2021Symmetry,Myers2021NJP,Myers2021PRXQ,Myers2022NJP,Myers2023Nanomat,Pena2023Entropy}.

In the present work, we focus on plasma engines that run in endorversible Stirling cycles. It is interesting to note that there are also several accounts in the literature of the fact that endoreversible Stirling cycles at maximum power operating with classical, ideal gases are described by $\eta_\mrm{CA}$ \cite{Erbay1997,Blanck1994Energy,Kaushik2000Energy}. Given that  it is often a good assumption (in first order approximation) that also plasmas can be described as ideal gases \cite{Slattery1980PRA} one is tempted to conclude that clearly also plasma engines at maximal power have the Curzon-Ahlborn efficiency. However, given that previous treatments of the endoreversible Stirling make, sometimes implicitly, often explicitly the assumption that the working medium is a regular, classical, ideal gas, it is not immediately obvious that plasmas do not require a separate treatment. 

Therefore, we start with a detailed discussion on the necessary conditions under which the Curzon-Ahlborn efficiency arises. Particular emphasis is placed on a comprehensive and pedagogical derivation, which then also leads to immediate generalizations. In fact, we will see that any gases that are described by caloric equations of states that are linear in temperature and at most additive in volume lead to the Curzon-Ahlborn efficiency. Neither the mechanical equation of state, nor the full fundamental relation is required.

This means, in particular, that also endoreversible Stirling engines whose working plasmas require a second-order virial expansion have the same efficiency. In the simplest case, their corresponding gas law is given by the van der Waals equation of state. 

As an example of plasma engines that fall not within this class, we then analyze an engine that operates with an electron-positron-photon plasma. For high enough temperatures \cite{Faussurier2024PP} such plasmas can be described by the photonic equation of state, which permits an almost completely analytical treatment. We find that photonic Stirling engines have a significantly smaller efficiency at maximal power. This is in stark contrast to Otto engines, in which case photonic working mediums lead to higher efficiency \cite{Smith2020JNET}.

\section{One-component plasmas -- modified ideal gas}

We start with a Stirling engine that operates with a one-component plasma. For such situations one commonly assumes that the plasma can be treated as an ``effective ideal gas'' \cite{Slattery1980PRA}. A classical, ideal gas is comprised of uniform, non-interacting, identical particles, whose caloric equation of state is proportional to temperature and independent of the volume,
\begin{equation}
E= \varsigma T\,.
\end{equation}
The constant $\varsigma$ depends on natural constants and the number of degrees of freedom. For instance, for a classical ideal gas in three spatial dimensions we have \cite{Callen1985}, $\varsigma=3/2 N\,k_B$, where $N$ is the number of particles and $k_B$ is Boltzmann's constant.

A one-component plasma is comprised of non-interacting, identical particles. At the fundamental level, a one-component plasma is nothing but a collection of uniform ions, that have been immersed in an equally and oppositely charged background, neutralizing the total charge of the plasma, and thus preventing any potential interparticle-Coulomb interactions \cite{Slattery1980PRA}. Thus, it follows that the corresponding caloric equation of state for the one-component plasma is still directly proportional to the temperature, and all ``quantum'' modifications can be accounted for by an effective number of degrees of freedom along with an arrangement of phenomenological constants that are unique to that plasma. We can write,
\begin{equation}
\label{eq:energy}
E=E_0+f\, T
\end{equation}
where $E_0$ is a constant off-setting the background energy, and $f$ quantifies the effective degrees of freedom \cite{Slattery1980PRA}. As we will see shortly, Eq.~\eqref{eq:energy} is a necessary and sufficient condition to obtain the Curzon-Ahlborn efficiency.

\subsection{Endoreversible cycle and efficiency}

In complete analogy to the endoreversible Carnot \cite{Curzon1975AJP}, Otto \cite{Deffner2018Entropy,Myers2020PRE,Smith2020JNET,Myers2021Symmetry,Myers2021NJP,Myers2021PRXQ,Myers2022NJP,Myers2023Nanomat,Pena2023Entropy}, and Brayton cycles \cite{Ferketic2023EPL}, we now construct the endoreversible Stirling cycle. 

\begin{figure}
    \centering
    \includegraphics[width=0.45\textwidth]{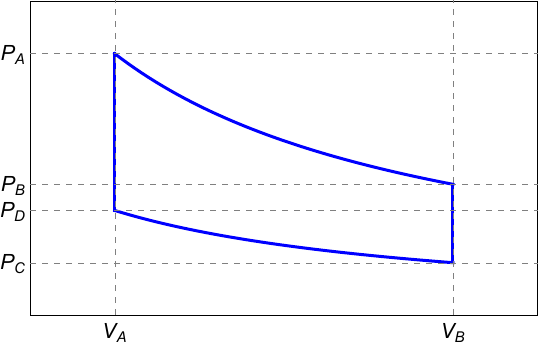}
    \hfill 
    \includegraphics[width=0.45\textwidth]{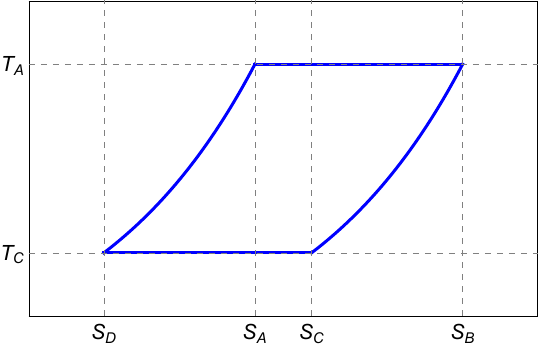}
    \caption{\label{fig:PV_TS} Schematic $PV$- and $TS$-diagrams of the Stirling cycle for the one-component plasma \eqref{eq:energy} as a working medium.}
\end{figure}

The Stirling cycle is a 4-stroke process comprising isothermal expansion, isochoric cooling, isothermal compression, and isochoric heating. The corresponding $PV$- and $TS$-diagrams for a one-component plasma \eqref{eq:energy} are depicted in Fig.~\ref{fig:PV_TS}. 

\subparagraph{$A\rightarrow B$: Isothermal expansion}

During the hot isotherm, the plasma is in contact with a heat reservoir at temperature $T_h$. However, as usual in endoreversible thermodynamics \cite{Hoffmann1997}, we assume that the plasma has not fully equilibrated with the hear reservoir, and rather has a temperature ``slightly'' below $T_h$, namely we have $T_{h,p}<T_h$.  Fourier's law then dictates that the heat flux is linear in the temperature gradient \cite{Curzon1975AJP}, and we can write
\begin{equation}
\label{eq:AB}
    Q_\mrm{A B}=\alpha\, t_{AB}\,\left(T_h-T_{h,p}\right)\,,
\end{equation}
where $\alpha$ is the thermal conductivity and $t_{AB}$ is the duration of the stroke.

Now note that for one-component plasmas the internal energy is constant for isothermal processes, cf. Eq,~\eqref{eq:energy}. Thus, we also immediately have,
\begin{equation}
\Delta E=E_B-E_A\quad\text{and}\quad W_{AB}=-Q_{AB},
\end{equation}
namely the work produced during the isothermal expansion is equal and opposite in sign to the heat absorbed from the heat reservoir.

\subparagraph{$B\rightarrow C$: Isochoric cooling}

During the isochoric stroke, the plasma is disconnected from the heat reservoirs. Thus, this stroke is identical to the ideal cycle. In any case, we have
\begin{equation}
W_{BC}=0\quad\text{and}\quad Q_{BC}=E_B-E_C=f\, \left(T_B-T_C\right)
\end{equation}
where we used the caloric equation of state \eqref{eq:energy}.

\subparagraph{$C\rightarrow D$: Isothermal compression}

In complete analogy to the hot isotherm, during the cold, isothermal compression the plasma is in contact with a cold reservoir at temperature $T_c$. However, we again assume that the plasma has not fully equilibrated with the reservoir, and that its temperature, $T_{c,p}$, is slightly above $T_c$. Again, employing Fourier's law, we can write
\begin{equation}
\label{eq:CD}
Q_{CD}=\beta \,t_{CD}\,\left(T_c-T_{c,p}\right)
\end{equation}
where $\beta$ is the thermal conductivity for the cold stroke, and $t_{CD}$ is the duration. As before, we can also write
\begin{equation}
\Delta E=E_D-E_C\quad\text{and}\quad W_{CD}=-Q_{CD}\,,
\end{equation}
which follows from the caloric equation of state \eqref{eq:energy}.

\subparagraph{$D\rightarrow A$: Isochoric heating}

The cycle is completed with another isochoric stroke. Again, the plasma is disconnected from the heat reservoirs. We have 
\begin{equation}
W_{DA}=0\quad\text{and}\quad Q_{DA}=f\, \left(T_D-T_A\right)
\end{equation}

\subparagraph{Endoreversible efficiency}

We are now interested in the efficiency at maximal power. To this end, consider that the plasma absorbs heat from the hot reservoir during the hot isotherm $A\rightarrow B$. Thus, we can write
\begin{equation}
\eta\equiv -\frac{W_\mrm{cyc}}{Q_{AB}}=1+\frac{Q_{CD}}{Q_{AB}}\,,
\end{equation}
where we used that the work produced during the entire cycle is $W_\mrm{cyc}=-(Q_{AB}+Q_{CD})$. Using Eqs.~\eqref{eq:AB} and \eqref{eq:CD} we also have,
\begin{equation}
\label{eq:eta}
\eta=1+\frac{\beta \,t_{CD}}{\alpha\, t_{AB}}\,\frac{ T_c-T_{c,p}}{T_h-T_{h,p}}\,,
\end{equation}
which appears to suggest that the efficiency depends on the stroke times.

However, the expression for the efficiency \eqref{eq:eta} can be further simplified by employing the entropy balance over one cycle. In fact, we have
\begin{equation}
0=\Delta S_\mrm{cyc}=\Delta S_{AB}+\Delta S_{BC}+\Delta S_{CD} +\Delta S_{DA}\,.
\end{equation}
Along the isothermal strokes, $A\rightarrow B$ and $C\rightarrow D$, the entropy is simply given by the heat divided by temperature,
\begin{equation}
\Delta S_{AB}=\frac{Q_{AB}}{T_{h,p}}=\frac{\alpha\, t_{AB}\,\left(T_h-T_{h,p}\right)}{T_{h,p}}
\end{equation}
and
\begin{equation}
 \Delta S_{CD}=\frac{Q_{CD}}{T_{c,p}}=\frac{\beta \,t_{CD}\,\left(T_c-T_{c,p}\right)}{T_{c,p}}\,,
\end{equation}
where we again used Eqs.~\eqref{eq:AB} and \eqref{eq:CD}.

For the isochoric strokes, we again exploit \emph{only} the caloric equation of state \eqref{eq:energy}. In its differential form, along the isochor, $\text{\dj}  W=0$, we have $dE=\text{\dj}Q=T\,dS.$ Thus, the total change in entropy can be written as
\begin{equation}
\label{eq:entropy_balance}
    \Delta S=\int_{S_i}^{S_f}\, dS=\int_{T_i}^{T_f}dT\, \frac{f}{T} \,=f\ln\left(\frac{T_f}{T_i}\right)\,.
\end{equation}
Consequently, we obtain
\begin{equation}
\Delta S_{BC}=-\Delta S_{DA}=f\,\ln\left(\frac{T_{c,p}}{T_{h,p}}\right)\,,
\end{equation}
and thus,
\begin{equation}
\label{eq:temps}
\frac{\alpha\, t_{AB}\,\left(T_h-T_{h,p}\right)}{T_{h,p}}=\frac{\beta \,t_{CD}\,\left(T_{c,p}-T_c\right)}{T_{c,p}}\,.
\end{equation}
In other words, the efficiency \eqref{eq:eta} of an endoreversible Stirling cycle operating with a one-component plasma simply becomes,
\begin{equation}
\label{eq:eta_stir}
\eta=1-\frac{T_{c,p}}{T_{h,p}}\,,
\end{equation}
which is identical to the ideal efficiency \cite{Callen1985} replacing the temperatures of the heat reservoirs with the corresponding temperatures of the plasma.

\subsection{Efficiency at maximal power}

As stated above, we are now interested in the efficiency at maximal power output. Inspecting Eq.~\eqref{eq:eta_stir} we need to to determine the temperatures, $T_{h,p}$ and $T_{c,p}$, that maximize the power. As usual, we write
\begin{equation}
\label{eq:pow}
P\equiv -\frac{W_\mrm{cyc}}{\tau_\mrm{cyc}}=\frac{Q_{AB}+Q_{CD}}{\gamma\, (t_{AB}+t_{CD})}\,,
\end{equation}
where $\gamma$ is a real constant. It will prove convenient to introduce the variables
\begin{equation}
x\equiv T_h-T_{h,p}\quad \text{and} \quad y\equiv T_{c,p}-T_c
\end{equation}
which is identical to the notation introduced by the original treatment by Curzon and Ahlborn \cite{Curzon1975AJP}. 

After a few lines of simple algebra, we obtain
\begin{equation}
\label{eq:power}
P(x,y)=\frac{\alpha\beta\,x y\,\left[(T_h-x)+(T_c+y)\right]}{\gamma \left[\alpha\,x\,(T_h-x)+\beta\,y\,(T_c+y)\right]}\,,
\end{equation}
where we once again employed Eqs.~\eqref{eq:AB} and \eqref{eq:CD}. The maximum of $P(x,y)$ is determine using standard calculus, namely solving $\pd_x P(x,y)=0$ and $\pd_y P(x,y)$ for $x$ and $y$.

Denoting the solutions by $x^*$ and $y^*$, we find,
\begin{equation}
x^*=\frac{\sqrt{\beta}}{\sqrt{\alpha}+\sqrt{\beta}}\,\left(T_h-\sqrt{T_h T_c}\right)
\end{equation}
and
\begin{equation}
y^*=\frac{\sqrt{\alpha}}{\sqrt{\alpha}+\sqrt{\beta}}\,\left(-T_c+\sqrt{T_h T_c}\right)\,.
\end{equation}
Substituting these solutions into the expression for the efficiency \eqref{eq:eta_stir} and simplifying the expression we finally obtain
\begin{equation}
\label{eq:Curzon}
\eta=1-\sqrt{\frac{T_c}{T_h}}\,.
\end{equation}
That is, the efficiency at maximal power of an endoreversible Stirling engine operating with a one-component plasma as working medium is given by the Curzon-Ahlborn efficiency. Our result \eqref{eq:Curzon} corroborates earlier findings for classical ideal gases \cite{Erbay1997,Blanck1994Energy,Kaushik2000Energy}. Thus, engines operating in endoreversible Stirling cycles with one-component plasmas and classical ideal gases have the same efficiency.

\subsection{Generalization to second order virial expansion}

Before we move on to more intricate working mediums, we emphasize that the present analysis is entirely based on the fact that the caloric equation of state is of the form of an ideal gas \eqref{eq:energy}. In particular, we did not need to require the mechanical equation of state to be given by the ideal gas law, nor did we need the full fundamental relation.

Interestingly, equations of state  of the form \eqref{eq:energy} are \emph{not} restricted to \emph{classical} ideal gases. For instance, also classical harmonic oscillators \cite{Deffner2018Entropy} and rubber bands \cite{Callen1985} are described by equations of state that are linear in temperature (and independent of volume).

For the present purposes, it is even more interesting to observe that the van der Waals gas is described by \cite{Callen1985}
\begin{equation}
\label{eq:vdW}
E=a_0\,T-\frac{b_0}{V^2}\,,
\end{equation}
where $a_0$ and $b_0$ are phenomenological constants. It is a simple exercise to show that the van der Waals equation of state is obtained from the virial expansion in second order \cite{Peliti2024}. Since the dependence on the volume $V$ in Eq.~\eqref{eq:vdW} is only additive, the changes in entropy during isochoric strokes  remain identical to above \eqref{eq:entropy_balance}. Consequently, we obtain the same expressions for efficiency \eqref{eq:eta_stir}, power \eqref{eq:power}, and efficiency at maximal power \eqref{eq:Curzon}.

It is worth emphasizing that for many real plasmas virial coefficients have been determined, see for instance \cite{Gervois1976PLA,Alastuey1992EPL,Omarbakiyeva2010PRE,Mattiello2010EPJC,Ropke2021PRE}. Our present results remain valid for any endoreversible Stirling cycle for plasma, for which a description of up to second order in the virial expansion is a good description.

\section{Relativistic electron-positron-photon plasma}

In the preceding section we analyzed the performance of endoreversible plasma engines that are described by equations of state which are (i) linear in temperature and (ii) at most additive in volume. The situation becomes more involved if these conditions are not met.

To this end, we conclude the analysis with another type of plasma, namely the relativistic electron–positron–photon plasma. Rather recently, it was shown in Ref.~\cite{Faussurier2024PP} that in the high-temperature limit the corresponding caloric equation of state can be written as the familiar photonic gas law \cite{Callen1985},
\begin{equation}
\label{eq:energy_photo}
E=\epsilon V T^4\,,
\end{equation}
where $\epsilon$ is a constant collecting natural constants, such as the speed of light, the electron mass, and $\hbar$, see Ref.~\cite{Faussurier2024PP}. In principle, we could also work with the more involved expression at finite temperature, but this would necessitate a fully numerical treatment. We choose to continue with Eq.~\eqref{eq:energy_photo} as it allows an almost entirely analytical treatment.

Note that Eq.~\eqref{eq:energy_photo} does not have the simple, linear form of Eq.~\eqref{eq:energy} that we exploited above. Therefore, the natural question arises whether engines with photonic plasmas operate at higher or lower efficiency than those with one-component plasmas. Note that for the same question, but for Otto cycles it was found that photonic gases significantly outperform classical ideal gases \cite{Smith2020JNET}.

\subsection{Endoreversible Stirling cycle}

\begin{figure}
    \centering
    \includegraphics[width=0.45\textwidth]{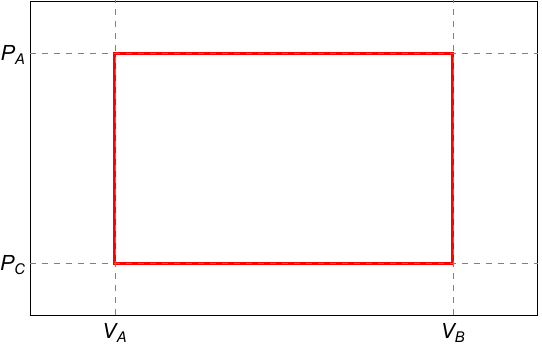}
    \hfill 
    \includegraphics[width=0.45\textwidth]{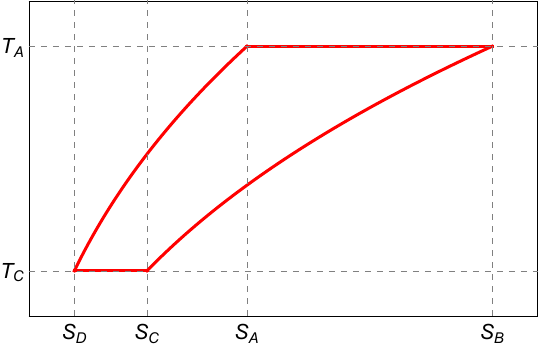}
    \caption{\label{fig:PV_TS_photo} Schematic $PV$- and $TS$-diagrams of the Stirling cycle for the photonic gas \eqref{eq:energy_photo} as a working medium.}
\end{figure}

To answer this question, we repeat the construction of the endoreversible Stirling cycle paying special attention to the required modifications. In Fig.~\ref{fig:PV_TS_photo} we plot the corresponding $PV$- and $TS$-diagrams schematically.

\subparagraph{$A\rightarrow B$: Isothermal expansion}

As before, we assume that during the hot isotherm the plasma is slightly colder than the hot heat reservoir, and that the heat flux is given by Fourier's law. Hence, we can write again $
Q_{AB}=\alpha t_{AB} x$. However, in this case we can no longer assume that the internal energy is constant, as Eq.~\eqref{eq:energy_photo} depends multiplicatively on the volume. Hence, we have to write for the work
\begin{equation}
\label{eq:work_AB}
W_{AB}=\Delta E_{AB}-Q_{AB}=\epsilon\, T_{h,p}^4\, (V_B-V_A)-\alpha t_{AB} x\,,
\end{equation}
which explicitly depends on the change in volume as well as the stroke time.

It is a standard exercise \cite{Callen1985} to show that the entropy of the photonic gas reads
\begin{equation}
\label{eq:entropy_photo}
S=\frac{4}{3}\epsilon V T^3\,.
\end{equation}
And, as discussed above, for isothermal processes, we simply have $Q_{AB}=T_{h,p}\,\Delta S_{AB}$. Therefore, we can also write
\begin{equation}
\label{eq:tAB}
\alpha t_{AB} x=\frac{4}{3}\, T_{h,p}^4\,(V_B-V_A)\,,
\end{equation}
which will become useful shortly.

\subparagraph{$B\rightarrow C$: Isochoric cooling}

For the isochoric strokes, we again have that $W_{BC}=0$ and 
\begin{equation}
Q_{BC}=\Delta E_{BC}= \epsilon V_B (T_{c,p}^4-T_{h,p}^4)\,. 
\end{equation}
However, as seen above, the work and heat of the isochoric strokes will not be required for further analysis.

\subparagraph{$C\rightarrow D$: Isothermal compression}

In complete analogy to the hot isotherm, we now can write $Q_{CD}=-\beta t_{CD} y$, and
\begin{equation}
\label{eq:work_CD}
W_{CD}=\Delta E_{CD}-Q_{CD}=- \epsilon\, T_{c,p}^4\, (V_B-V_A)-Q_{CD}\,.
\end{equation}
Moreover, again employing the expression for the entropy \eqref{eq:entropy_photo} and $Q_{CD}=T_{c,p}\, \Delta S_{CD}$ we also have
\begin{equation}
\label{eq:tCD}
\beta t_{CD} y=\frac{4}{3}\, T_{c,p}^4\,(V_B-V_A)\,.
\end{equation}

\subparagraph{$D\rightarrow A$: Isochoric heating}

For completeness we also collect the work, $W_{DA}=0$, and heat,
\begin{equation}
Q_{DA}=\Delta E_{DA}=\epsilon V_A (T_{h,p}^4-T_{c,p}^4)\,,
\end{equation}
during the isochoric heating stroke.

\subparagraph{Endoreversible efficiency}

\begin{figure}
    \centering
    \includegraphics[width=0.45\textwidth]{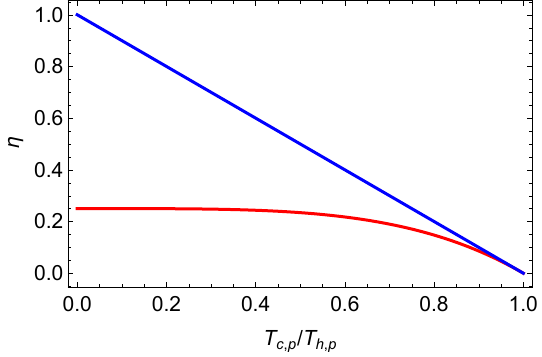}
    \caption{Endoreversible Stirling efficiency  for the one-component plasma \eqref{eq:Curzon} (blue line) and the photonic gas  \eqref{eq:eta_photo} (red line).}
    \label{fig:efficiency}
\end{figure}

Before we start analyzing the power, we again first derive an expression for the efficiency,
\begin{equation}
\label{eq:eta_phot_def}
\eta\equiv -\frac{W_\mrm{cyc}}{Q_{AB}}=-\frac{W_{AB}+W_{CD}}{Q_{AB}}\,,
\end{equation}
Substituting Eqs.~\eqref{eq:work_AB}, \eqref{eq:tAB}, \eqref{eq:work_CD}, and \eqref{eq:tCD} into the definition \eqref{eq:eta_phot_def} we obtain,
\begin{equation}
\label{eq:eta_photo}
\eta=\frac{1}{4}\left[1-\left(\frac{T_{c,p}}{T_{h,p}}\right)^4\right]\,.
\end{equation}
Comparing the latter result \eqref{eq:eta_photo} with the endoreversible efficiency for one-component plasmas \eqref{eq:eta_stir} we note that endoreversible Stirling efficiency for photonic gases is always smaller than Eq.~\eqref{eq:eta_stir}, cf. Fig.~\ref{fig:efficiency}.

\subsection{Efficiency at maximal power}

The obvious question now is what this means for the efficiency at maximal power. To this end, we again consider the power output,
\begin{equation}
P=\frac{Q_{AB}+Q_{CD}}{\gamma\, (t_{AB}+t_{CD})}\,,
\end{equation}
which in the present case can be written as
\begin{equation}
\label{eq:power_photo}
P(T_{c,p},T_{h,p})=\frac{\alpha\beta}{4 \gamma}\frac{(T_h-T_{h,p})(T_c-T_{c,p})(T_{h,p}^4-T_{c,p}^4)}{\beta T_{h,p}^4(T_c-T_{c,p})-\alpha T_{c,p}^4(T_h-T_{h,p})}\,.
\end{equation}
We immediately observe that Eq.~\eqref{eq:power_photo} is significantly more involved than the power for one-component plasmas \eqref{eq:power}, and thus we have to resort to a numerical analysis.

\begin{figure}
    \centering
    \includegraphics[width=0.45\textwidth]{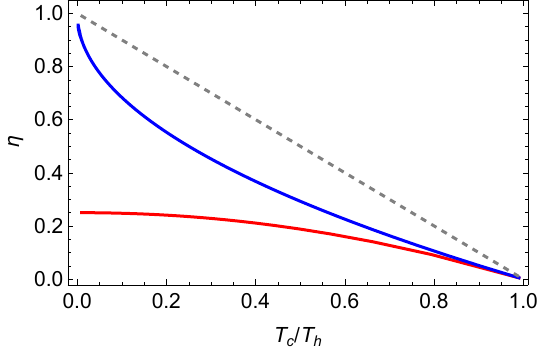}
    \caption{Efficiency at maximal power for the photonic equation of state \eqref{eq:energy_photo} (red line) together with the Curzon-Ahlborn efficiency 
    \eqref{eq:Curzon} (blue line), and the Carnot efficiency, $\eta_C=1-T_c/T_h$, (gray, dashed line). Parameters are $\alpha=1$, $\beta=1$, and $\gamma=1$.}
    \label{fig:efficiency_max}
\end{figure}

In Fig.~\ref{fig:efficiency_max} we depict the solution for one set of parameters. We observe that also at maximal power, the efficiency for photonic gases is significantly below the Curzon-Ahlborn efficiency \eqref{eq:Curzon}. We emphasize again that this is in contrast to endoreversible Otto cycle, in which the photonic case has a higher efficiency \cite{Smith2020JNET}.

\section{Concluding remarks}

In the present work, we analyzed the efficiency at maximal power of endoreversible Stirling cycles. As working mediums, we considered one-component plasmas with caloric equations of state, extensions in second order virial expansion, and photonic gases. We found that for all thermodynamics systems, whose caloric equation of state only depends linearly on temperature and at most additively on volume, the efficiency at maximal power is given by the seminal Curzon-Ahlborn efficiency. Interestingly, the efficiency for photonic working mediums is significantly smaller, which is in stark contrast to endoreversible Otto cycles.

Particular emphasis was put on a comprehensive, self-contained, and pedagogical presentation of the treatment. While our work is purely theoretical and it is unlikely that our results will lead directly to the experimental realization of a practically useful plasma Stirling engine, the conceptual mathematical steps of our analysis could be applied to more realistic scenarios. 
The only necessary ``ingredient'' is the caloric equation of state for any considered plasma, which then step-for-step leads to the corresponding expression for the efficiency at maximal power.

\acknowledgements{S.D. acknowledges support from the John Templeton Foundation under Grant No. 62422.}

\appendix 

\section{Equal strokes times -- trivial case}

In this appendix, we briefly discuss a trivial case for the one-component plasma \eqref{eq:energy}. As done previously for Otto \cite{Deffner2018Entropy} and Brayton \cite{Ferketic2023EPL} cycles, one is tempted to make a ``simplifying'' assumption. Namely, to set the stroke times equal to each other, $\tau \equiv t_{AB}=t_{CD}$. In this case, we can Eq.~\eqref{eq:pow} as
\begin{equation}
P=\frac{1}{2 \gamma}\,(\alpha\,x-\beta \,y)\,,
\end{equation}
which is linear in $x$ and $y$. In this case, we can determine the maximal power simply from the maximal and minimal values of $x$ and $y$, respectively. Recognizing that at no instant the plasma can be colder than the cold reservoir, we can write
\begin{equation}
x_\mrm{max}=T_h-T_c\quad \text{and}\quad y_\mrm{min}=0\,.
\end{equation}
Substituting the latter into Eq.~\eqref{eq:eta} we obtain
\begin{equation}
\eta=1-\frac{\beta}{\alpha} \frac{y_\mrm{min}}{x_\mrm{max}}=1\,.
\end{equation}

\bibliography{plasma_bib}

\begin{thebibliography}{37}%
\makeatletter
\providecommand \@ifxundefined [1]{%
 \@ifx{#1\undefined}
}%
\providecommand \@ifnum [1]{%
 \ifnum #1\expandafter \@firstoftwo
 \else \expandafter \@secondoftwo
 \fi
}%
\providecommand \@ifx [1]{%
 \ifx #1\expandafter \@firstoftwo
 \else \expandafter \@secondoftwo
 \fi
}%
\providecommand \natexlab [1]{#1}%
\providecommand \enquote  [1]{``#1''}%
\providecommand \bibnamefont  [1]{#1}%
\providecommand \bibfnamefont [1]{#1}%
\providecommand \citenamefont [1]{#1}%
\providecommand \href@noop [0]{\@secondoftwo}%
\providecommand \href [0]{\begingroup \@sanitize@url \@href}%
\providecommand \@href[1]{\@@startlink{#1}\@@href}%
\providecommand \@@href[1]{\endgroup#1\@@endlink}%
\providecommand \@sanitize@url [0]{\catcode `\\12\catcode `\$12\catcode
  `\&12\catcode `\#12\catcode `\^12\catcode `\_12\catcode `\%12\relax}%
\providecommand \@@startlink[1]{}%
\providecommand \@@endlink[0]{}%
\providecommand \url  [0]{\begingroup\@sanitize@url \@url }%
\providecommand \@url [1]{\endgroup\@href {#1}{\urlprefix }}%
\providecommand \urlprefix  [0]{URL }%
\providecommand \Eprint [0]{\href }%
\providecommand \doibase [0]{https://doi.org/}%
\providecommand \selectlanguage [0]{\@gobble}%
\providecommand \bibinfo  [0]{\@secondoftwo}%
\providecommand \bibfield  [0]{\@secondoftwo}%
\providecommand \translation [1]{[#1]}%
\providecommand \BibitemOpen [0]{}%
\providecommand \bibitemStop [0]{}%
\providecommand \bibitemNoStop [0]{.\EOS\space}%
\providecommand \EOS [0]{\spacefactor3000\relax}%
\providecommand \BibitemShut  [1]{\csname bibitem#1\endcsname}%
\let\auto@bib@innerbib\@empty
\bibitem [{\citenamefont {Conde}(2020)}]{Conde2020}%
  \BibitemOpen
  \bibfield  {author} {\bibinfo {author} {\bibfnamefont {L.}~\bibnamefont
  {Conde}},\ }\href {https://doi.org/10.1088/978-0-7503-3543-0} {\emph
  {\bibinfo {title} {An Introduction to Plasma Physics and its Space
  Applications, Volume 2}}},\ 2053-2563\ (\bibinfo  {publisher} {IOP
  Publishing},\ \bibinfo {year} {2020})\BibitemShut {NoStop}%
\bibitem [{\citenamefont {Morse}(2018)}]{morse2018nuclear}%
  \BibitemOpen
  \bibfield  {author} {\bibinfo {author} {\bibfnamefont {E.}~\bibnamefont
  {Morse}},\ }\href@noop {} {\emph {\bibinfo {title} {Nuclear fusion}}}\
  (\bibinfo  {publisher} {Springer},\ \bibinfo {year} {2018})\BibitemShut
  {NoStop}%
\bibitem [{\citenamefont {Department~of Energy}()}]{fusion}%
  \BibitemOpen
  \bibfield  {author} {\bibinfo {author} {\bibfnamefont {U.~S.~A.}\
  \bibnamefont {Department~of Energy}},\ }\href
  {https://www.energy.gov/science/fes/fusion-energy-sciences} {\bibinfo {title}
  {Fusion energy sciences}}\BibitemShut {NoStop}%
\bibitem [{\citenamefont {Segantin}\ \emph {et~al.}(2020)\citenamefont
  {Segantin}, \citenamefont {Bersano}, \citenamefont {Falcone},\ and\
  \citenamefont {Testoni}}]{Segantin2020FED}%
  \BibitemOpen
  \bibfield  {author} {\bibinfo {author} {\bibfnamefont {S.}~\bibnamefont
  {Segantin}}, \bibinfo {author} {\bibfnamefont {A.}~\bibnamefont {Bersano}},
  \bibinfo {author} {\bibfnamefont {N.}~\bibnamefont {Falcone}},\ and\ \bibinfo
  {author} {\bibfnamefont {R.}~\bibnamefont {Testoni}},\ }\bibfield  {title}
  {\bibinfo {title} {Exploration of power conversion thermodynamic cycles for
  arc fusion reactor},\ }\href
  {https://doi.org/https://doi.org/10.1016/j.fusengdes.2020.111645} {\bibfield
  {journal} {\bibinfo  {journal} {Fusion Engineering and Design}\ }\textbf
  {\bibinfo {volume} {155}},\ \bibinfo {pages} {111645} (\bibinfo {year}
  {2020})}\BibitemShut {NoStop}%
\bibitem [{\citenamefont {Terahara}\ and\ \citenamefont
  {Tanabe}(2023)}]{Terahara2023FED}%
  \BibitemOpen
  \bibfield  {author} {\bibinfo {author} {\bibfnamefont {Y.}~\bibnamefont
  {Terahara}}\ and\ \bibinfo {author} {\bibfnamefont {K.}~\bibnamefont
  {Tanabe}},\ }\bibfield  {title} {\bibinfo {title} {Process design of a
  thermochemical cycle for hydrogen production compatible with nuclear fusion
  heat sources},\ }\href
  {https://doi.org/https://doi.org/10.1016/j.fusengdes.2023.113868} {\bibfield
  {journal} {\bibinfo  {journal} {Fusion Engineering and Design}\ }\textbf
  {\bibinfo {volume} {194}},\ \bibinfo {pages} {113868} (\bibinfo {year}
  {2023})}\BibitemShut {NoStop}%
\bibitem [{\citenamefont {Avinash}(2010)}]{Avinash2010PP}%
  \BibitemOpen
  \bibfield  {author} {\bibinfo {author} {\bibfnamefont {K.}~\bibnamefont
  {Avinash}},\ }\bibfield  {title} {\bibinfo {title} {Plasma heat pump and heat
  engine},\ }\href {https://doi.org/10.1063/1.3467034} {\bibfield  {journal}
  {\bibinfo  {journal} {Phys. Plasmas}\ }\textbf {\bibinfo {volume} {17}},\
  \bibinfo {pages} {082105} (\bibinfo {year} {2010})}\BibitemShut {NoStop}%
\bibitem [{\citenamefont {Avinash}\ and\ \citenamefont
  {Chaudhary}(2014)}]{Avinash2014PINSA}%
  \BibitemOpen
  \bibfield  {author} {\bibinfo {author} {\bibfnamefont {K.}~\bibnamefont
  {Avinash}}\ and\ \bibinfo {author} {\bibfnamefont {S.}~\bibnamefont
  {Chaudhary}},\ }\bibfield  {title} {\bibinfo {title} {Stirling like engine
  using plasma electric fields},\ }\href
  {https://doi.org/10.16943/ptinsa/2014/v80i5/47976} {\bibfield  {journal}
  {\bibinfo  {journal} {Proc. Indian Natn. Sci. Acad.}\ }\textbf {\bibinfo
  {volume} {80}},\ \bibinfo {pages} {1099} (\bibinfo {year}
  {2014})}\BibitemShut {NoStop}%
\bibitem [{\citenamefont {Hoffmann}\ \emph {et~al.}(1997)\citenamefont
  {Hoffmann}, \citenamefont {Burzler},\ and\ \citenamefont
  {Schubert}}]{Hoffmann1997}%
  \BibitemOpen
  \bibfield  {author} {\bibinfo {author} {\bibfnamefont {K.~H.}\ \bibnamefont
  {Hoffmann}}, \bibinfo {author} {\bibfnamefont {J.~M.}\ \bibnamefont
  {Burzler}},\ and\ \bibinfo {author} {\bibfnamefont {S.}~\bibnamefont
  {Schubert}},\ }\bibfield  {title} {\bibinfo {title} {Endoreversible
  thermodynamics},\ }\href {https://doi.org/10.1515/jnet.1997.22.4.311}
  {\bibfield  {journal} {\bibinfo  {journal} {J. Non-Equilib. Thermodyn.}\
  }\textbf {\bibinfo {volume} {22}},\ \bibinfo {pages} {311} (\bibinfo {year}
  {1997})}\BibitemShut {NoStop}%
\bibitem [{\citenamefont {Curzon}\ and\ \citenamefont
  {Ahlborn}(1975)}]{Curzon1975AJP}%
  \BibitemOpen
  \bibfield  {author} {\bibinfo {author} {\bibfnamefont {F.~L.}\ \bibnamefont
  {Curzon}}\ and\ \bibinfo {author} {\bibfnamefont {B.}~\bibnamefont
  {Ahlborn}},\ }\bibfield  {title} {\bibinfo {title} {{Efficiency of a {Carnot}
  engine at maximum power output}},\ }\href {https://doi.org/10.1119/1.10023}
  {\bibfield  {journal} {\bibinfo  {journal} {Am. J. Phys.}\ }\textbf {\bibinfo
  {volume} {43}},\ \bibinfo {pages} {22} (\bibinfo {year} {1975})}\BibitemShut
  {NoStop}%
\bibitem [{\citenamefont {Deffner}(2018)}]{Deffner2018Entropy}%
  \BibitemOpen
  \bibfield  {author} {\bibinfo {author} {\bibfnamefont {S.}~\bibnamefont
  {Deffner}},\ }\bibfield  {title} {\bibinfo {title} {Efficiency of harmonic
  quantum otto engines at maximal power},\ }\bibfield  {journal} {\bibinfo
  {journal} {Entropy}\ }\textbf {\bibinfo {volume} {20}},\ \href
  {https://doi.org/10.3390/e20110875} {10.3390/e20110875} (\bibinfo {year}
  {2018})\BibitemShut {NoStop}%
\bibitem [{\citenamefont {Ferketic}\ and\ \citenamefont
  {Deffner}(2023)}]{Ferketic2023EPL}%
  \BibitemOpen
  \bibfield  {author} {\bibinfo {author} {\bibfnamefont {E.~E.}\ \bibnamefont
  {Ferketic}}\ and\ \bibinfo {author} {\bibfnamefont {S.}~\bibnamefont
  {Deffner}},\ }\bibfield  {title} {\bibinfo {title} {Boosting thermodynamic
  performance by bending space-time},\ }\href
  {https://doi.org/10.1209/0295-5075/acad9c} {\bibfield  {journal} {\bibinfo
  {journal} {EPL (Europhys. Lett.)}\ }\textbf {\bibinfo {volume} {141}},\
  \bibinfo {pages} {19001} (\bibinfo {year} {2023})}\BibitemShut {NoStop}%
\bibitem [{\citenamefont {Esposito}\ \emph {et~al.}(2010)\citenamefont
  {Esposito}, \citenamefont {Kawai}, \citenamefont {Lindenberg},\ and\
  \citenamefont {Van~den Broeck}}]{Esposito2010}%
  \BibitemOpen
  \bibfield  {author} {\bibinfo {author} {\bibfnamefont {M.}~\bibnamefont
  {Esposito}}, \bibinfo {author} {\bibfnamefont {R.}~\bibnamefont {Kawai}},
  \bibinfo {author} {\bibfnamefont {K.}~\bibnamefont {Lindenberg}},\ and\
  \bibinfo {author} {\bibfnamefont {C.}~\bibnamefont {Van~den Broeck}},\
  }\bibfield  {title} {\bibinfo {title} {Efficiency at maximum power of
  low-dissipation {Carnot} engines},\ }\href
  {https://doi.org/10.1103/PhysRevLett.105.150603} {\bibfield  {journal}
  {\bibinfo  {journal} {Phys. Rev. Lett.}\ }\textbf {\bibinfo {volume} {105}},\
  \bibinfo {pages} {150603} (\bibinfo {year} {2010})}\BibitemShut {NoStop}%
\bibitem [{\citenamefont {Leff}(1987)}]{Leff1987}%
  \BibitemOpen
  \bibfield  {author} {\bibinfo {author} {\bibfnamefont {H.~S.}\ \bibnamefont
  {Leff}},\ }\bibfield  {title} {\bibinfo {title} {Thermal efficiency at
  maximum work output: {New} results for old heat engines},\ }\href
  {https://doi.org/10.1119/1.15071} {\bibfield  {journal} {\bibinfo  {journal}
  {Am. J. Phys.}\ }\textbf {\bibinfo {volume} {55}},\ \bibinfo {pages} {602}
  (\bibinfo {year} {1987})}\BibitemShut {NoStop}%
\bibitem [{\citenamefont {Rezek}\ and\ \citenamefont
  {Kosloff}(2006)}]{Rezek2006}%
  \BibitemOpen
  \bibfield  {author} {\bibinfo {author} {\bibfnamefont {Y.}~\bibnamefont
  {Rezek}}\ and\ \bibinfo {author} {\bibfnamefont {R.}~\bibnamefont
  {Kosloff}},\ }\bibfield  {title} {\bibinfo {title} {Irreversible performance
  of a quantum harmonic heat engine},\ }\href
  {https://doi.org/10.1088/1367-2630/8/5/083} {\bibfield  {journal} {\bibinfo
  {journal} {New J. Phys.}\ }\textbf {\bibinfo {volume} {8}},\ \bibinfo {pages}
  {83} (\bibinfo {year} {2006})}\BibitemShut {NoStop}%
\bibitem [{\citenamefont {Abah}\ \emph {et~al.}(2012)\citenamefont {Abah},
  \citenamefont {Ro\ss{}nagel}, \citenamefont {Jacob}, \citenamefont {Deffner},
  \citenamefont {Schmidt-Kaler}, \citenamefont {Singer},\ and\ \citenamefont
  {Lutz}}]{Abah2012}%
  \BibitemOpen
  \bibfield  {author} {\bibinfo {author} {\bibfnamefont {O.}~\bibnamefont
  {Abah}}, \bibinfo {author} {\bibfnamefont {J.}~\bibnamefont {Ro\ss{}nagel}},
  \bibinfo {author} {\bibfnamefont {G.}~\bibnamefont {Jacob}}, \bibinfo
  {author} {\bibfnamefont {S.}~\bibnamefont {Deffner}}, \bibinfo {author}
  {\bibfnamefont {F.}~\bibnamefont {Schmidt-Kaler}}, \bibinfo {author}
  {\bibfnamefont {K.}~\bibnamefont {Singer}},\ and\ \bibinfo {author}
  {\bibfnamefont {E.}~\bibnamefont {Lutz}},\ }\bibfield  {title} {\bibinfo
  {title} {Single-ion heat engine at maximum power},\ }\href
  {https://doi.org/10.1103/PhysRevLett.109.203006} {\bibfield  {journal}
  {\bibinfo  {journal} {Phys. Rev. Lett.}\ }\textbf {\bibinfo {volume} {109}},\
  \bibinfo {pages} {203006} (\bibinfo {year} {2012})}\BibitemShut {NoStop}%
\bibitem [{\citenamefont {Deffner}\ and\ \citenamefont
  {Campbell}(2019)}]{Deffner2019book}%
  \BibitemOpen
  \bibfield  {author} {\bibinfo {author} {\bibfnamefont {S.}~\bibnamefont
  {Deffner}}\ and\ \bibinfo {author} {\bibfnamefont {S.}~\bibnamefont
  {Campbell}},\ }\href {https://doi.org/10.1088/2053-2571/ab21c6} {\emph
  {\bibinfo {title} {Quantum Thermodynamics}}}\ (\bibinfo  {publisher} {Morgan
  \& Claypool Publishers},\ \bibinfo {year} {2019})\BibitemShut {NoStop}%
\bibitem [{\citenamefont {Kloc}\ \emph {et~al.}(2019)\citenamefont {Kloc},
  \citenamefont {Cejnar},\ and\ \citenamefont {Schaller}}]{Kloc2019}%
  \BibitemOpen
  \bibfield  {author} {\bibinfo {author} {\bibfnamefont {M.}~\bibnamefont
  {Kloc}}, \bibinfo {author} {\bibfnamefont {P.}~\bibnamefont {Cejnar}},\ and\
  \bibinfo {author} {\bibfnamefont {G.}~\bibnamefont {Schaller}},\ }\bibfield
  {title} {\bibinfo {title} {Collective performance of a finite-time quantum
  otto cycle},\ }\href {https://doi.org/10.1103/PhysRevE.100.042126} {\bibfield
   {journal} {\bibinfo  {journal} {Phys. Rev. E}\ }\textbf {\bibinfo {volume}
  {100}},\ \bibinfo {pages} {042126} (\bibinfo {year} {2019})}\BibitemShut
  {NoStop}%
\bibitem [{\citenamefont {Myers}\ and\ \citenamefont
  {Deffner}(2020)}]{Myers2020PRE}%
  \BibitemOpen
  \bibfield  {author} {\bibinfo {author} {\bibfnamefont {N.~M.}\ \bibnamefont
  {Myers}}\ and\ \bibinfo {author} {\bibfnamefont {S.}~\bibnamefont
  {Deffner}},\ }\bibfield  {title} {\bibinfo {title} {Bosons outperform
  fermions: The thermodynamic advantage of symmetry},\ }\href
  {https://doi.org/10.1103/PhysRevE.101.012110} {\bibfield  {journal} {\bibinfo
   {journal} {Phys. Rev. E}\ }\textbf {\bibinfo {volume} {101}},\ \bibinfo
  {pages} {012110} (\bibinfo {year} {2020})}\BibitemShut {NoStop}%
\bibitem [{\citenamefont {Smith}\ \emph {et~al.}(2020)\citenamefont {Smith},
  \citenamefont {Pal},\ and\ \citenamefont {Deffner}}]{Smith2020JNET}%
  \BibitemOpen
  \bibfield  {author} {\bibinfo {author} {\bibfnamefont {Z.}~\bibnamefont
  {Smith}}, \bibinfo {author} {\bibfnamefont {P.~S.}\ \bibnamefont {Pal}},\
  and\ \bibinfo {author} {\bibfnamefont {S.}~\bibnamefont {Deffner}},\
  }\bibfield  {title} {\bibinfo {title} {Endoreversible otto engines at maximal
  power},\ }\href {https://doi.org/doi:10.1515/jnet-2020-0039} {\bibfield
  {journal} {\bibinfo  {journal} {Journal of Non-Equilibrium Thermodynamics}\
  }\textbf {\bibinfo {volume} {45}},\ \bibinfo {pages} {305} (\bibinfo {year}
  {2020})}\BibitemShut {NoStop}%
\bibitem [{\citenamefont {Myers}\ \emph
  {et~al.}(2021{\natexlab{a}})\citenamefont {Myers}, \citenamefont {McCready},\
  and\ \citenamefont {Deffner}}]{Myers2021Symmetry}%
  \BibitemOpen
  \bibfield  {author} {\bibinfo {author} {\bibfnamefont {N.~M.}\ \bibnamefont
  {Myers}}, \bibinfo {author} {\bibfnamefont {J.}~\bibnamefont {McCready}},\
  and\ \bibinfo {author} {\bibfnamefont {S.}~\bibnamefont {Deffner}},\
  }\bibfield  {title} {\bibinfo {title} {Quantum heat engines with singular
  interactions},\ }\bibfield  {journal} {\bibinfo  {journal} {Symmetry}\
  }\textbf {\bibinfo {volume} {13}},\ \href
  {https://doi.org/10.3390/sym13060978} {10.3390/sym13060978} (\bibinfo {year}
  {2021}{\natexlab{a}})\BibitemShut {NoStop}%
\bibitem [{\citenamefont {Myers}\ \emph
  {et~al.}(2021{\natexlab{b}})\citenamefont {Myers}, \citenamefont {Abah},\
  and\ \citenamefont {Deffner}}]{Myers2021NJP}%
  \BibitemOpen
  \bibfield  {author} {\bibinfo {author} {\bibfnamefont {N.~M.}\ \bibnamefont
  {Myers}}, \bibinfo {author} {\bibfnamefont {O.}~\bibnamefont {Abah}},\ and\
  \bibinfo {author} {\bibfnamefont {S.}~\bibnamefont {Deffner}},\ }\bibfield
  {title} {\bibinfo {title} {Quantum otto engines at relativistic energies},\
  }\href {https://doi.org/10.1088/1367-2630/ac2756} {\bibfield  {journal}
  {\bibinfo  {journal} {New J. Phys.}\ }\textbf {\bibinfo {volume} {23}},\
  \bibinfo {pages} {105001} (\bibinfo {year} {2021}{\natexlab{b}})}\BibitemShut
  {NoStop}%
\bibitem [{\citenamefont {Myers}\ and\ \citenamefont
  {Deffner}(2021)}]{Myers2021PRXQ}%
  \BibitemOpen
  \bibfield  {author} {\bibinfo {author} {\bibfnamefont {N.~M.}\ \bibnamefont
  {Myers}}\ and\ \bibinfo {author} {\bibfnamefont {S.}~\bibnamefont
  {Deffner}},\ }\bibfield  {title} {\bibinfo {title} {Thermodynamics of
  statistical anyons},\ }\href {https://doi.org/10.1103/PRXQuantum.2.040312}
  {\bibfield  {journal} {\bibinfo  {journal} {PRX Quantum}\ }\textbf {\bibinfo
  {volume} {2}},\ \bibinfo {pages} {040312} (\bibinfo {year}
  {2021})}\BibitemShut {NoStop}%
\bibitem [{\citenamefont {Myers}\ \emph {et~al.}(2022)\citenamefont {Myers},
  \citenamefont {Peña}, \citenamefont {Negrete}, \citenamefont {Vargas},
  \citenamefont {De~Chiara},\ and\ \citenamefont {Deffner}}]{Myers2022NJP}%
  \BibitemOpen
  \bibfield  {author} {\bibinfo {author} {\bibfnamefont {N.~M.}\ \bibnamefont
  {Myers}}, \bibinfo {author} {\bibfnamefont {F.~J.}\ \bibnamefont {Peña}},
  \bibinfo {author} {\bibfnamefont {O.}~\bibnamefont {Negrete}}, \bibinfo
  {author} {\bibfnamefont {P.}~\bibnamefont {Vargas}}, \bibinfo {author}
  {\bibfnamefont {G.}~\bibnamefont {De~Chiara}},\ and\ \bibinfo {author}
  {\bibfnamefont {S.}~\bibnamefont {Deffner}},\ }\bibfield  {title} {\bibinfo
  {title} {Boosting engine performance with bose–einstein condensation},\
  }\href {https://doi.org/10.1088/1367-2630/ac47cc} {\bibfield  {journal}
  {\bibinfo  {journal} {New J. Phys.}\ }\textbf {\bibinfo {volume} {24}},\
  \bibinfo {pages} {025001} (\bibinfo {year} {2022})}\BibitemShut {NoStop}%
\bibitem [{\citenamefont {Myers}\ \emph {et~al.}(2023)\citenamefont {Myers},
  \citenamefont {Peña}, \citenamefont {Cortés},\ and\ \citenamefont
  {Vargas}}]{Myers2023Nanomat}%
  \BibitemOpen
  \bibfield  {author} {\bibinfo {author} {\bibfnamefont {N.~M.}\ \bibnamefont
  {Myers}}, \bibinfo {author} {\bibfnamefont {F.~J.}\ \bibnamefont {Peña}},
  \bibinfo {author} {\bibfnamefont {N.}~\bibnamefont {Cortés}},\ and\ \bibinfo
  {author} {\bibfnamefont {P.}~\bibnamefont {Vargas}},\ }\bibfield  {title}
  {\bibinfo {title} {Multilayer graphene as an endoreversible otto engine},\
  }\bibfield  {journal} {\bibinfo  {journal} {Nanomaterials}\ }\textbf
  {\bibinfo {volume} {13}},\ \href {https://doi.org/10.3390/nano13091548}
  {10.3390/nano13091548} (\bibinfo {year} {2023})\BibitemShut {NoStop}%
\bibitem [{\citenamefont {Peña}\ \emph {et~al.}(2023)\citenamefont {Peña},
  \citenamefont {Myers}, \citenamefont {Órdenes}, \citenamefont
  {Albarrán-Arriagada},\ and\ \citenamefont {Vargas}}]{Pena2023Entropy}%
  \BibitemOpen
  \bibfield  {author} {\bibinfo {author} {\bibfnamefont {F.~J.}\ \bibnamefont
  {Peña}}, \bibinfo {author} {\bibfnamefont {N.~M.}\ \bibnamefont {Myers}},
  \bibinfo {author} {\bibfnamefont {D.}~\bibnamefont {Órdenes}}, \bibinfo
  {author} {\bibfnamefont {F.}~\bibnamefont {Albarrán-Arriagada}},\ and\
  \bibinfo {author} {\bibfnamefont {P.}~\bibnamefont {Vargas}},\ }\bibfield
  {title} {\bibinfo {title} {Enhanced efficiency at maximum power in a
  fock–darwin model quantum dot engine},\ }\bibfield  {journal} {\bibinfo
  {journal} {Entropy}\ }\textbf {\bibinfo {volume} {25}},\ \href
  {https://doi.org/10.3390/e25030518} {10.3390/e25030518} (\bibinfo {year}
  {2023})\BibitemShut {NoStop}%
\bibitem [{\citenamefont {Erbay}\ and\ \citenamefont
  {Yavuz}(1997)}]{Erbay1997}%
  \BibitemOpen
  \bibfield  {author} {\bibinfo {author} {\bibfnamefont {L.~B.}\ \bibnamefont
  {Erbay}}\ and\ \bibinfo {author} {\bibfnamefont {H.}~\bibnamefont {Yavuz}},\
  }\bibfield  {title} {\bibinfo {title} {Analysis of the {Stirling} heat engine
  at maximum power conditions},\ }\href
  {https://doi.org/10.1016/S0360-5442(96)00159-4} {\bibfield  {journal}
  {\bibinfo  {journal} {Energy}\ }\textbf {\bibinfo {volume} {22}},\ \bibinfo
  {pages} {645} (\bibinfo {year} {1997})}\BibitemShut {NoStop}%
\bibitem [{\citenamefont {Blank}\ \emph {et~al.}(1994)\citenamefont {Blank},
  \citenamefont {Davis},\ and\ \citenamefont {Wu}}]{Blanck1994Energy}%
  \BibitemOpen
  \bibfield  {author} {\bibinfo {author} {\bibfnamefont {D.~A.}\ \bibnamefont
  {Blank}}, \bibinfo {author} {\bibfnamefont {G.~W.}\ \bibnamefont {Davis}},\
  and\ \bibinfo {author} {\bibfnamefont {C.}~\bibnamefont {Wu}},\ }\bibfield
  {title} {\bibinfo {title} {Power optimization of an endoreversible stirling
  cycle with regeneration},\ }\href
  {https://doi.org/https://doi.org/10.1016/0360-5442(94)90111-2} {\bibfield
  {journal} {\bibinfo  {journal} {Energy}\ }\textbf {\bibinfo {volume} {19}},\
  \bibinfo {pages} {125} (\bibinfo {year} {1994})}\BibitemShut {NoStop}%
\bibitem [{\citenamefont {Kaushik}\ and\ \citenamefont
  {Kumar}(2000)}]{Kaushik2000Energy}%
  \BibitemOpen
  \bibfield  {author} {\bibinfo {author} {\bibfnamefont {S.}~\bibnamefont
  {Kaushik}}\ and\ \bibinfo {author} {\bibfnamefont {S.}~\bibnamefont
  {Kumar}},\ }\bibfield  {title} {\bibinfo {title} {Finite time thermodynamic
  analysis of endoreversible stirling heat engine with regenerative losses},\
  }\href {https://doi.org/https://doi.org/10.1016/S0360-5442(00)00023-2}
  {\bibfield  {journal} {\bibinfo  {journal} {Energy}\ }\textbf {\bibinfo
  {volume} {25}},\ \bibinfo {pages} {989} (\bibinfo {year} {2000})}\BibitemShut
  {NoStop}%
\bibitem [{\citenamefont {Slattery}\ \emph {et~al.}(1980)\citenamefont
  {Slattery}, \citenamefont {Doolen},\ and\ \citenamefont
  {DeWitt}}]{Slattery1980PRA}%
  \BibitemOpen
  \bibfield  {author} {\bibinfo {author} {\bibfnamefont {W.~L.}\ \bibnamefont
  {Slattery}}, \bibinfo {author} {\bibfnamefont {G.~D.}\ \bibnamefont
  {Doolen}},\ and\ \bibinfo {author} {\bibfnamefont {H.~E.}\ \bibnamefont
  {DeWitt}},\ }\bibfield  {title} {\bibinfo {title} {Improved equation of state
  for the classical one-component plasma},\ }\href
  {https://doi.org/10.1103/PhysRevA.21.2087} {\bibfield  {journal} {\bibinfo
  {journal} {Phys. Rev. A}\ }\textbf {\bibinfo {volume} {21}},\ \bibinfo
  {pages} {2087} (\bibinfo {year} {1980})}\BibitemShut {NoStop}%
\bibitem [{\citenamefont {Faussurier}(2024)}]{Faussurier2024PP}%
  \BibitemOpen
  \bibfield  {author} {\bibinfo {author} {\bibfnamefont {G.}~\bibnamefont
  {Faussurier}},\ }\bibfield  {title} {\bibinfo {title} {Equation of state of
  the relativistic electron–positron–photon plasma at arbitrary temperature
  and degeneracy},\ }\href {https://doi.org/10.1063/5.0230675} {\bibfield
  {journal} {\bibinfo  {journal} {Phys. Plasmas}\ }\textbf {\bibinfo {volume}
  {31}},\ \bibinfo {pages} {102702} (\bibinfo {year} {2024})}\BibitemShut
  {NoStop}%
\bibitem [{\citenamefont {Callen}(1985)}]{Callen1985}%
  \BibitemOpen
  \bibfield  {author} {\bibinfo {author} {\bibfnamefont {H.}~\bibnamefont
  {Callen}},\ }\href@noop {} {\emph {\bibinfo {title} {Thermodynamics and an
  Introduction to Thermostastistics}}}\ (\bibinfo  {publisher} {Wiley},\
  \bibinfo {address} {New York, USA},\ \bibinfo {year} {1985})\BibitemShut
  {NoStop}%
\bibitem [{\citenamefont {Peliti}(2024)}]{Peliti2024}%
  \BibitemOpen
  \bibfield  {author} {\bibinfo {author} {\bibfnamefont {L.}~\bibnamefont
  {Peliti}},\ }\href@noop {} {\emph {\bibinfo {title} {Statistical mechanics in
  a nutshell}}},\ \bibinfo {edition} {2nd}\ ed.\ (\bibinfo  {publisher}
  {Princeton University Press},\ \bibinfo {year} {2024})\BibitemShut {NoStop}%
\bibitem [{\citenamefont {Gervois}\ \emph {et~al.}(1976)\citenamefont
  {Gervois}, \citenamefont {Pomeau},\ and\ \citenamefont
  {Résibois}}]{Gervois1976PLA}%
  \BibitemOpen
  \bibfield  {author} {\bibinfo {author} {\bibfnamefont {A.}~\bibnamefont
  {Gervois}}, \bibinfo {author} {\bibfnamefont {Y.}~\bibnamefont {Pomeau}},\
  and\ \bibinfo {author} {\bibfnamefont {P.}~\bibnamefont {Résibois}},\
  }\bibfield  {title} {\bibinfo {title} {Non equilibrium density expansion in
  the dilute one component plasma},\ }\href
  {https://doi.org/https://doi.org/10.1016/0375-9601(76)90697-6} {\bibfield
  {journal} {\bibinfo  {journal} {Phys. Lett. A}\ }\textbf {\bibinfo {volume}
  {55}},\ \bibinfo {pages} {343} (\bibinfo {year} {1976})}\BibitemShut
  {NoStop}%
\bibitem [{\citenamefont {Alastuey}\ and\ \citenamefont
  {Perez}(1992)}]{Alastuey1992EPL}%
  \BibitemOpen
  \bibfield  {author} {\bibinfo {author} {\bibfnamefont {A.}~\bibnamefont
  {Alastuey}}\ and\ \bibinfo {author} {\bibfnamefont {A.}~\bibnamefont
  {Perez}},\ }\bibfield  {title} {\bibinfo {title} {Virial expansion of the
  equation of state of a quantum plasma},\ }\href
  {https://doi.org/10.1209/0295-5075/20/1/004} {\bibfield  {journal} {\bibinfo
  {journal} {EPL (Europhys. Lett.)}\ }\textbf {\bibinfo {volume} {20}},\
  \bibinfo {pages} {19} (\bibinfo {year} {1992})}\BibitemShut {NoStop}%
\bibitem [{\citenamefont {Omarbakiyeva}\ \emph {et~al.}(2010)\citenamefont
  {Omarbakiyeva}, \citenamefont {Fortmann}, \citenamefont {Ramazanov},\ and\
  \citenamefont {R\"opke}}]{Omarbakiyeva2010PRE}%
  \BibitemOpen
  \bibfield  {author} {\bibinfo {author} {\bibfnamefont {Y.~A.}\ \bibnamefont
  {Omarbakiyeva}}, \bibinfo {author} {\bibfnamefont {C.}~\bibnamefont
  {Fortmann}}, \bibinfo {author} {\bibfnamefont {T.~S.}\ \bibnamefont
  {Ramazanov}},\ and\ \bibinfo {author} {\bibfnamefont {G.}~\bibnamefont
  {R\"opke}},\ }\bibfield  {title} {\bibinfo {title} {Cluster virial expansion
  for the equation of state of partially ionized hydrogen plasma},\ }\href
  {https://doi.org/10.1103/PhysRevE.82.026407} {\bibfield  {journal} {\bibinfo
  {journal} {Phys. Rev. E}\ }\textbf {\bibinfo {volume} {82}},\ \bibinfo
  {pages} {026407} (\bibinfo {year} {2010})}\BibitemShut {NoStop}%
\bibitem [{\citenamefont {Mattiello}\ and\ \citenamefont
  {Cassing}(2010)}]{Mattiello2010EPJC}%
  \BibitemOpen
  \bibfield  {author} {\bibinfo {author} {\bibfnamefont {S.}~\bibnamefont
  {Mattiello}}\ and\ \bibinfo {author} {\bibfnamefont {W.}~\bibnamefont
  {Cassing}},\ }\bibfield  {title} {\bibinfo {title} {Shear viscosity of the
  quark--gluon plasma from a virial expansion},\ }\href
  {https://doi.org/10.1140/epjc/s10052-010-1459-3} {\bibfield  {journal}
  {\bibinfo  {journal} {Eur. Phys. J. C}\ }\textbf {\bibinfo {volume} {70}},\
  \bibinfo {pages} {243} (\bibinfo {year} {2010})}\BibitemShut {NoStop}%
\bibitem [{\citenamefont {R\"opke}\ \emph {et~al.}(2021)\citenamefont
  {R\"opke}, \citenamefont {Sch\"orner}, \citenamefont {Redmer},\ and\
  \citenamefont {Bethkenhagen}}]{Ropke2021PRE}%
  \BibitemOpen
  \bibfield  {author} {\bibinfo {author} {\bibfnamefont {G.}~\bibnamefont
  {R\"opke}}, \bibinfo {author} {\bibfnamefont {M.}~\bibnamefont {Sch\"orner}},
  \bibinfo {author} {\bibfnamefont {R.}~\bibnamefont {Redmer}},\ and\ \bibinfo
  {author} {\bibfnamefont {M.}~\bibnamefont {Bethkenhagen}},\ }\bibfield
  {title} {\bibinfo {title} {Virial expansion of the electrical conductivity of
  hydrogen plasmas},\ }\href {https://doi.org/10.1103/PhysRevE.104.045204}
  {\bibfield  {journal} {\bibinfo  {journal} {Phys. Rev. E}\ }\textbf {\bibinfo
  {volume} {104}},\ \bibinfo {pages} {045204} (\bibinfo {year}
  {2021})}\BibitemShut {NoStop}%
\end{thebibliography}%

\end{document}